\documentclass[aps,epsfigure,twocolumn,superscriptaddress,showkeys,nofootinbib]{revtex4-1}
\usepackage[colorlinks=true,linkcolor=blue,urlcolor=blue,citecolor=blue,pdfusetitle]{hyperref}
\usepackage[utf8]{inputenc}
\usepackage[english]{babel}
\usepackage{amsmath}
\usepackage[caption = false]{subfig}
\usepackage{graphicx,epstopdf}
\usepackage{blindtext}

\usepackage[table,xcdraw]{xcolor}
\usepackage{lipsum}
\usepackage{amsfonts}
\usepackage{bbm}
\usepackage{amssymb}
\usepackage{enumerate}
\usepackage{color}
\usepackage{latexsym}
\usepackage{times,txfonts}
\usepackage[normalem]{ulem}
\graphicspath{ {./figures/} }
\usepackage{amsmath}
\usepackage{tikz}
\usetikzlibrary{quantikz}
\usepackage{algpseudocode}
\usepackage{algorithm}
\begin{document}

\title{Quantum algorithm for finding minimum values in a Quantum Random Access Memory}
\author{Anton S. Albino}
\email{anton.albino@fieb.org.br}
\affiliation{Latin American Quantum Computing Center, SENAI CIMATEC, Salvador, Brazil.}

\author{Lucas Q. Galvão}
\email{lucas.g5847@ufob.edu.br}
\affiliation{Latin American Quantum Computing Center, SENAI CIMATEC, Salvador, Brazil.}

\author{Ethan Hansen}
\email{1ethanhansen@proton.me}
\affiliation{Zapata Computing, Canada}

\author{Mauro Q. Nooblath Neto}
\email{mauro.neto@fbter.org.br}
\affiliation{Latin American Quantum Computing Center, SENAI CIMATEC, Salvador, Brazil.}

\author{Clebson Cruz}
\email{clebson.cruz@ufob.edu.br}
\affiliation{Grupo de Informação Quântica, Centro de Ciências Exatas e das Tecnologias, Universidade Federal do Oeste da Bahia - Campus Reitor Edgard Santos. Rua Bertioga, 892, Morada Nobre I, 47810-059 Barreiras, Bahia, Brasil.}



\begin{abstract}
Finding the minimum value in an unordered database is a common and fundamental task in computer science. However, the optimal classical deterministic algorithm can find the minimum value with a time complexity that grows linearly with the number of elements in the database. In this paper, we present the proposal of a quantum algorithm for finding the minimum value of a database, which is quadratically faster than its best classical analogs. We assume a Quantum Random Access Memory (QRAM) that stores values from a database and perform an iterative search based on an oracle whose role is to limit the searched values by controlling the states of the most significant qubits. A complexity analysis was performed in order to demonstrate the advantage of this quantum algorithm over its classical counterparts. 
Furthermore, we demonstrate how the proposed algorithm would be used in an unsupervised machine learning task through a quantum version of the K-means algorithm.
\end{abstract}
\maketitle

\section{Introduction}

Random Access Memory (RAM) is a versatile, short-term memory used in computing for storing and retrieving information via bits \cite{sedra2004micro}. Similarly, the concept of Quantum RAM (QRAM) emerges with the same goal but employing qubits to apply a superposition of states to achieve faster results for computational applications, whether quantum or classical~\cite{giovannetti2008quantum,giovannetti2008architectures,park2019circuit,yuan2022optimal,phalak2022approximate,asaka2021quantum,de2020circuit,hann2021resilience,di2020fault,blencowe2010quantum}. Several works discuss the potential of its applications to optimize the execution of quantum algorithms, including quantum searching on a classical database \cite{hur2022quantum,di2020fault,broda2016quantum,giovannetti2008quantum2,lu2013efficient}, collision finding \cite{hosoyamada2020quantum,hur2022quantum,naya2019new,bonnetain2022finding}, and algorithms for solving linear systems \cite{duan2020survey,wossnig2018quantum,kerenidis2020quantum,shao2020row}, for instance.

These results have attracted the attention of the scientific community in the past few years, leading to the development of QRAM architectures that demonstrate the potential for producing efficient results in quantum computing \cite{giovannetti2008architectures,phalak2022approximate,hann2021resilience,asaka2021quantum,arunachalam2015robustness,bugalho2022resource,park2019circuit,paler2020parallelizing}. Some models, such as Fanout quantum RAM \cite{giovannetti2008architectures,hann2021resilience,asaka2021quantum} and Bucket-Brigade quantum RAM \cite{giovannetti2008architectures,arunachalam2015robustness,paler2020parallelizing},  illustrate potential future implementations of a QRAM in practical scenarios. In addition, recent experiments have revealed a designed architecture for hybrid quantum computers that use superconducting qubits and spin-qubit memory crystals capable, in theory, of implementing a QRAM in real systems \cite{blencowe2010quantum}.

Furthermore, other efforts have been made to construct quantum algorithms that are able to optimally access a QRAM in the process of searching for certain values stored in its cells \cite{di2020fault}. In general, problems based on searching use the famous Grover's algorithm to search quantum states in an unstructured list \cite{nielsen2002quantum,figgatt2017complete,hosoyamada2020quantum,seidel2023automatic,szablowski2021understanding}. On the other hand, a well-known example of determining the minimal value in a list is the so-called Dürr-Hoyer minimum finding algorithm \cite{durr1996quantum}, which employs Grover’s Algorithm as a fundamental subroutine to find the greatest or smallest entry in a list \cite{wiebe2015quantum}. 

In this scenario, based on the core concept of Durr-Hoyer's algorithm, we apply Grover's Algorithm as a subroutine to develop a quantum algorithm for identifying the smallest value in a classical data set stored in a QRAM. The proposed Quantum Minimum Search (QMS) algorithm is based on the iterative change of the oracle function, which limits the searched values by controlling the states of the most significant qubits. 
First, we present the description of the QMS algorithm, describing the concept of QRAM and approaching an example to find the minimum in a list of four real values using the proposed algorithm. In sequence, we analyze the complexity of the QMS algorithm compared with classical algorithms. The results show that, whereas the complexity of the classical algorithm grows linearly with the number of elements in the database, $O(N)$, since the classical algorithms go through all the $N$ items in the list, the presented QMS algorithm has a complexity of $O(\sqrt{\frac{N}{t}})$, with 
$t$ being the number of marked states. Finally, we present an application of the proposed algorithm 
in the K-means problem of determining the optimal location of K-centroids in order to minimize the sum of all distances between the points and their respective centroids.

\section{Quantum Minimum Search (QMS) Algorithm}

The search problem is a ubiquitous subject of discussion in classical computer science \cite{nielsen2002quantum}. 
The problem consists of identifying the index of the database item ($x$) that fulfills some predetermined search criterion $x = y$, where $y$ is the sought element, given an unstructured database with $N$ elements. In this context, it is possible to prepare the so-called response function ($R(x)$) that translates database entries to \texttt{True} if the entry $x$ matches the search criterion ($x = y$) or \texttt{False} if $x \neq y$. This is possible by using the so-called Oracle subroutine, which queries the database until the desired item is located. Consequently, the bigger the requested element's position in the list, the greater the number of queries required to locate it. Therefore, the complexity of this task is exactly proportional to the number of items on the list \cite{nielsen2002quantum,figgatt2017complete}. On average, $\frac{N}{2}$ queries are required, and the complexity of the classical search problem is thus defined as being of order $\mathcal{O}\left(N\right)$ \cite{szablowski2021understanding}.

The renowned quantum search algorithm developed by Grover searches unstructured datasets and comprises an application that demonstrates the advantages of quantum computing over classical analogs \cite{nielsen2002quantum}. The introduction of the quantum superposition concept enables the algorithm the ability to map all the database items simultaneously, which allows for a reduction in the total number of queries, which gives an improvement in the efficiency of the search process \cite{figgatt2017complete}. In this regard, Grover's Algorithm presents a complexity that grows in order of $O(\sqrt{N})$, being quadratically faster than its classical counterpart \cite{szablowski2021understanding}. Therefore, many algorithms use Grover's method as a subroutine in order to optimize some quantum processes. A famous example is the so-called  D\"{u}rr-Hoyer's Algorithm for finding a minimum value in an unstructured database \cite{chen2020low}. Based on the Grovers algorithm, the achieve a quadratic speed-up in the minimum search problem, which complexity can be expressed as $O(\sqrt{\frac{N}{t}})$, where $t$ is the number of marked states \cite{durr1996quantum}.

In this context, this study investigates the quantum minimum search (QMS) problem, proposing a quantum algorithm quadratically faster than any classical analogs for finding minimal values in a quantum random access memory (QRAM). We proposed the algorithm's description considering a data set represented by the vector $\vec{y}$, which can be rewritten in the computational bases as quantum states. The problem is to find the minimum value of the list, $y_{min}$, using Grover’s Algorithm as a subroutine based on Durr-Hoyer's approach.



\subsection{Quantum Random Access Memory (QRAM)}

In order to use Grover's Algorithm to find a minimum in a classical data set, one proposal is to use a QRAM, typically meaning a large classical memory, which can be queried in a quantum superposition. It can be built using an equivalent quantum circuit in which classical data is stored in a quantum register in binary form. It can be done by creating two quantum registers (with $n$ and $m$ qubits, respectively) whose initialization should be
\begin{equation}
|\psi_0\rangle = \frac{1}{\sqrt{2^n}}\sum_{x=0}^{2^n-1}|x\rangle\otimes |0\rangle^{\otimes m}~, \end{equation}
which can be implemented by the operation $H^{\otimes n}\otimes I^{\otimes m}$ that creates an equal superposition on the first register and keeps the second register in the state $|0\rangle^{\otimes m}$. The quantum RAM is implemented in the second register by applying an operator $U_X$ given by multicontrolled-NOT operations. The goal is to store the classical values $\vec{y} = \{y_0, y_1, y_2, ... , y_k\}$ into quantum states in order to obtain

\begin{equation}
|\psi_1\rangle = \frac{1}{\sqrt{2^n}}\sum_{x=0}^{2^n-1}|x\rangle\otimes |y_x\rangle.
\end{equation}

Thus, a quantum RAM can store $2^n$ data values. In this scenario, we need to choose the number $m$ of qubits used in the second register. Since we are searching for the minimum value of the whole dataset, a random index can serve as our first iteration. Thus, it is possible to specify the number $m$ such that the number associated with such an index can be expressed on a computational binary basis.

\subsection{Finding the minimum in a QRAM}

\begin{figure}[H]
    \centering
    \includegraphics[width=0.4\textwidth]{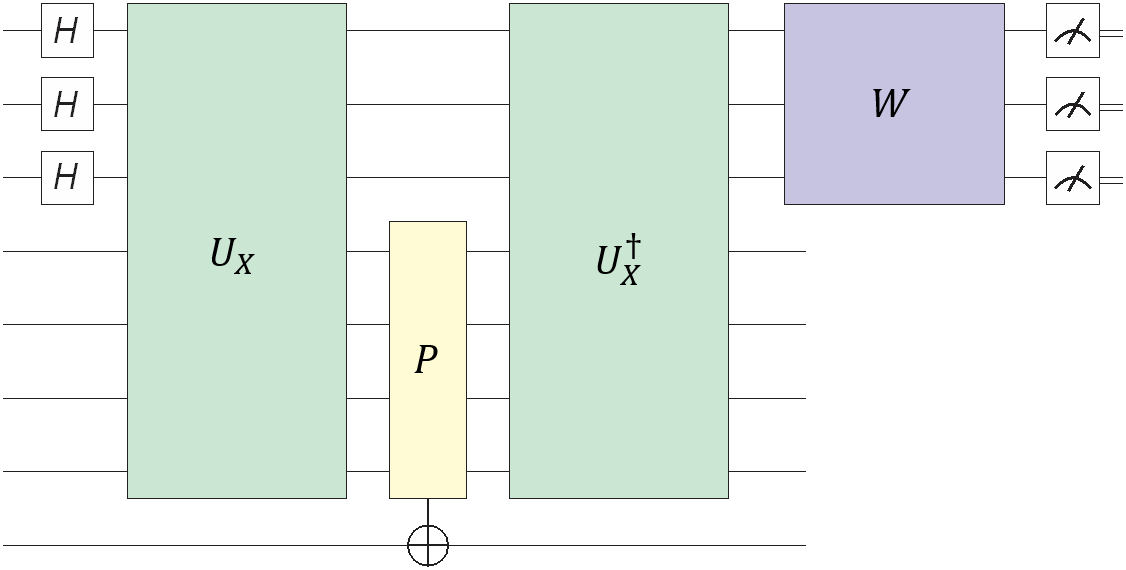}
    \caption{Full quantum circuit for minimum search.  $U_X$ is the representation of a QRAM; the operator $P$ is changed iteratively by analysing the most significant qubits in the last measurement; and $W$ is the diffuser operator. 
It is important to emphasize that the last qubit has its state initialized in $|-\rangle$.}
    \label{fullcirc}

\end{figure}

In order to perform the task of finding a smallest value stored in a QRAM, the adopted strategy is by performing a search analyzing the most (or less, if we want the maximum value) significant bits from a single measurement. The full quantum circuit can be seen in Fig. \ref{fullcirc}, where the special subroutine responsible for searching according to most significant qubits is the iterative phase flip, given by an operator $P$.

The key idea of the algorithm is in the dynamics of the $P$ operator. The additional register (qubits further down in Fig. \ref{fullcirc}) is used to represent the storage of classical values in QRAM and is also where the search is done. It is known that if the most significant qubits have bits in the $0$ state, it means that, in the decimal base, this number is smaller than if the most significant qubits were in the $1$ state. Based on this logic, the $P$ operator can be constructed through multicontrolled-NOT having qubits either with control at $0$ or with control at $1$. Therefore, the algorithm that governs the dynamics of $P$ is described on box Algorithm \texttt{1}.



\begin{algorithm}[H]
    \caption{Finding the minimum in a QRAM} 
    \begin{itemize}
        \item \textbf{Input} A classical database $\vec{y}$
        \begin{enumerate}
            \item Take a random value $y_{i} = f(x_i)$, whose binary representation demands $m$ bits. 
       
            \item Initialize a quantum computer in the state $|\psi_0\rangle = \frac{1}{\sqrt{2^n}} \sum_{x=0}^{2^n}|x\rangle|0\rangle^{\otimes m}|-\rangle$.
       
            \item Store the classical values in the QRAM in order to get the state \text{$|\psi_1\rangle = \frac{1}{\sqrt{2^n}}\sum_{x=0}^{2^n}|x\rangle|y_x\rangle|-\rangle$}
        
            \item Apply the oracle operator $P$ in order to guarantee that the most signicant qubit is \text{$0$}, that is, the marked states is less than \text{$y_{i}$}.
        
            \item Apply the diffuser operator, $W$, to amplify the marked states.
        
            \item Perform a measurement in the computational basis to obtain $y_{i+1} < y_{i}$.
        
            \item If all qubits have analyzed:
        
            \begin{itemize}
                \item \textbf{end if}
            \end{itemize}
            else:
            \begin{itemize}
                \item \textbf{repeat} steps
            \end{itemize}

        \end{enumerate}
        
        \item \textbf{return} $y_{i}$
    \end{itemize}
    
\end{algorithm}

Thus, Grover's Algorithm can be used iteratively in order to amplify states (index) that correspond to smaller values than the last one, quadratically faster than their classical counterparts. For instance, supposing the following dataset $\Vec{y} = \{5,4,12,10,8\}$, the list entries can be represented in the computational basis (with four qubits) as $\Vec{y} = \{|0101\rangle, |0100\rangle, |1100\rangle, |1010\rangle, |1000\rangle \}$.

\begin{figure} [H]
    \centering
    \includegraphics[width=0.4\textwidth]{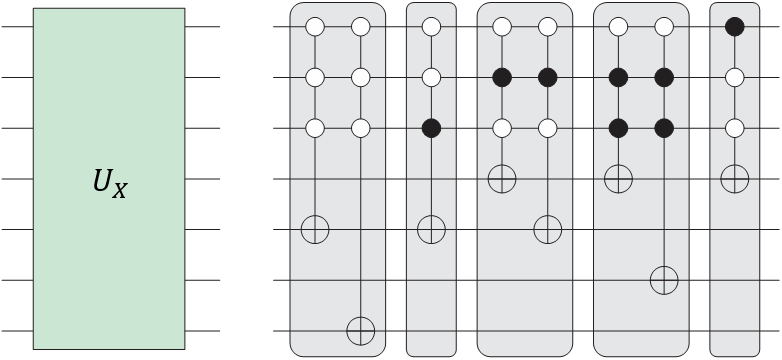}
    \caption{Implementation of a quantum RAM, given by the operator $U_X$, as a quantum circuit. Each gray block stores a classical value from the database.}
    \label{uxcircuit}

\end{figure}
If the first guess is (purely classical), for instance, $10 \rightarrow 1010$ it is very unlikely that this number is the lowest. This can be confirmed by looking for the number whose most significant qubit is $|0\rangle$.

\begin{figure} [H]
\centering
    \includegraphics[width=0.2\textwidth]{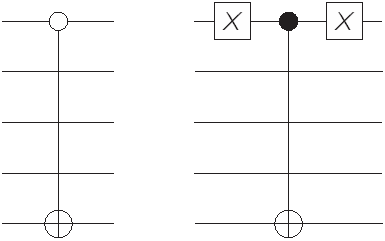}
    \caption{Quantum operator $P$ for searching all states whose the most significant qubit is in the state $|0\rangle$.}

\end{figure}

\begin{figure} [H]
\centering
    \includegraphics[width=0.35\textwidth]{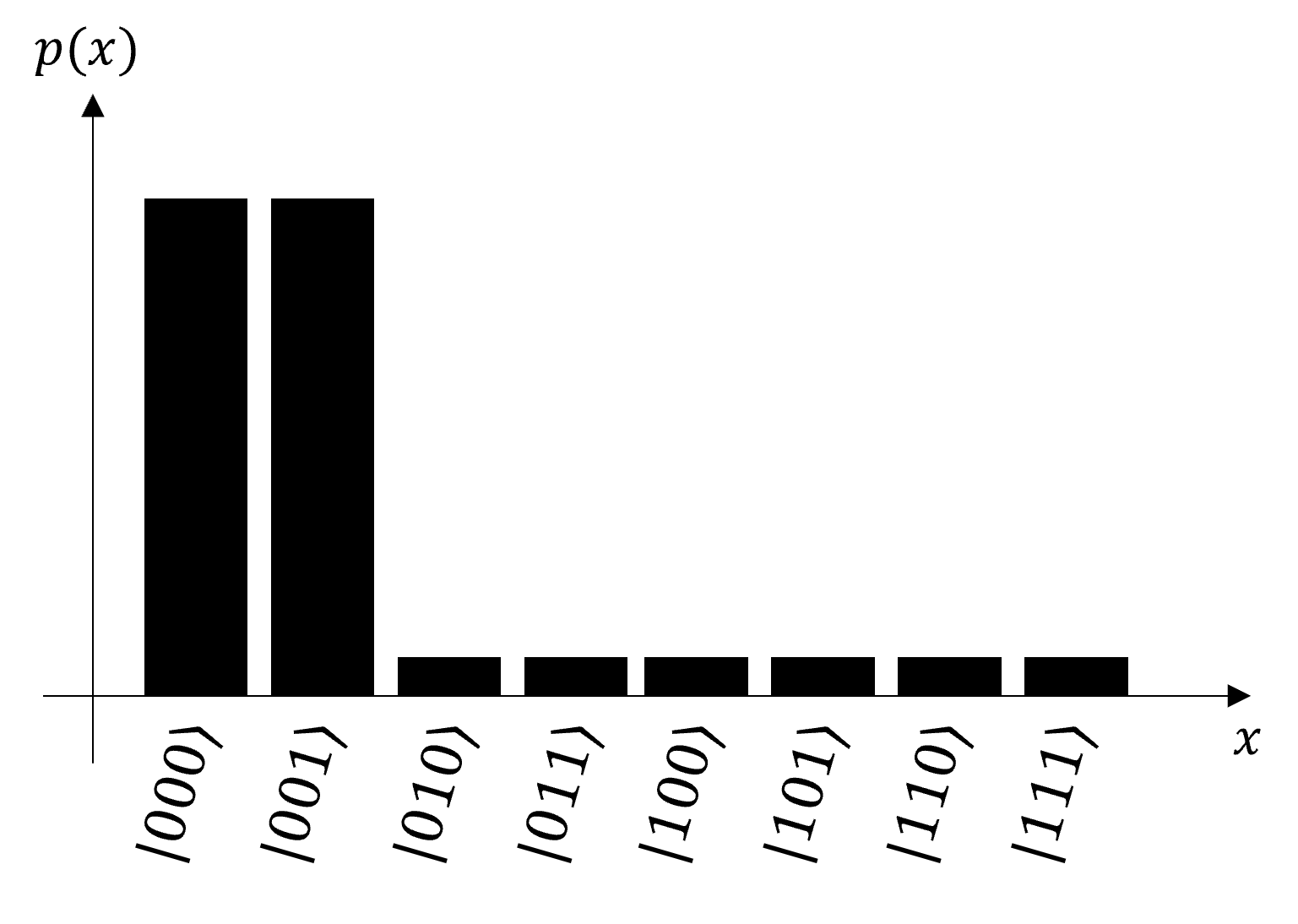}
    \caption{A diagramatic representation of the quantum state probabilities. The states $|x_0\rangle\rangle = |000\rangle$ and $|x_1\rangle = |001\rangle$ are amplified beacause $f(x_0) = |0101\rangle$ and $f(x_1) = |0100\rangle$ whose the most significant qubits are $|0\rangle$ for both.}
    \label{diag1}

\end{figure}

Thus, a Grover iteration with this oracle mark all states whose the most significant qubit is in state $|0\rangle$. A diagram representation of the state before the measurement can be seen in Fig. \ref{diag1}.

After that, by performing Grover's search in that most significant qubit, the states $|0100\rangle$ and $|0101\rangle$ will be had equal probability to be measured. If we get the state $|0101\rangle$ after the measurement, the next step is to search for values whose two first binary digits are $|00\rangle$. 

\begin{figure} [H]
    \centering
    \includegraphics[width=0.2\textwidth]{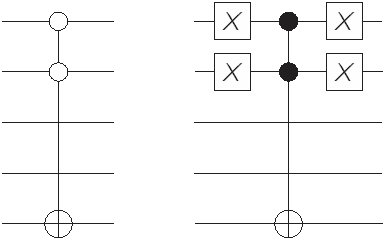}
    \caption{Quantum operator $P$ for searching all states whose the two most significant qubits are in the state $|00\rangle$.}

\end{figure}

If there are one or more with which it is satisfied, a number less than $|0101\rangle$ will be measured ($y_i > y_{i+1}$), if not, a number greater will likely be measured ($y_i < y_{i+1}$), because none rotation is performed in the initial state. 

In the case where $y_i < y_{i+1}$, the process shows that the minimum is $|0101\rangle$ or a less number whose the first two more significant qubits are also $|01\rangle$, so it is necessary to search for values whose third most significant qubits are $|010\rangle$. In this particular case, the only remaining task is to verify if $|0100\rangle$ is in the QRAM since it is the smaller possible number whose three most significant qubits are in the state $|010\rangle$.

\begin{figure} [H]
  \centering  \includegraphics[width=0.2\textwidth]{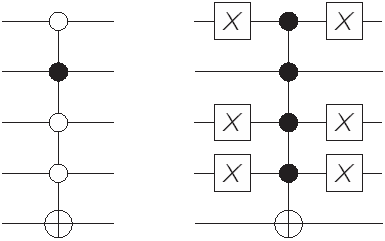}
    \caption{Quantum operator $P$ for checking if the value $|4\rangle \equiv |0100\rangle$ is in the QRAM.}

\end{figure}

\begin{figure} [H]
  \centering  \includegraphics[width=0.35\textwidth]{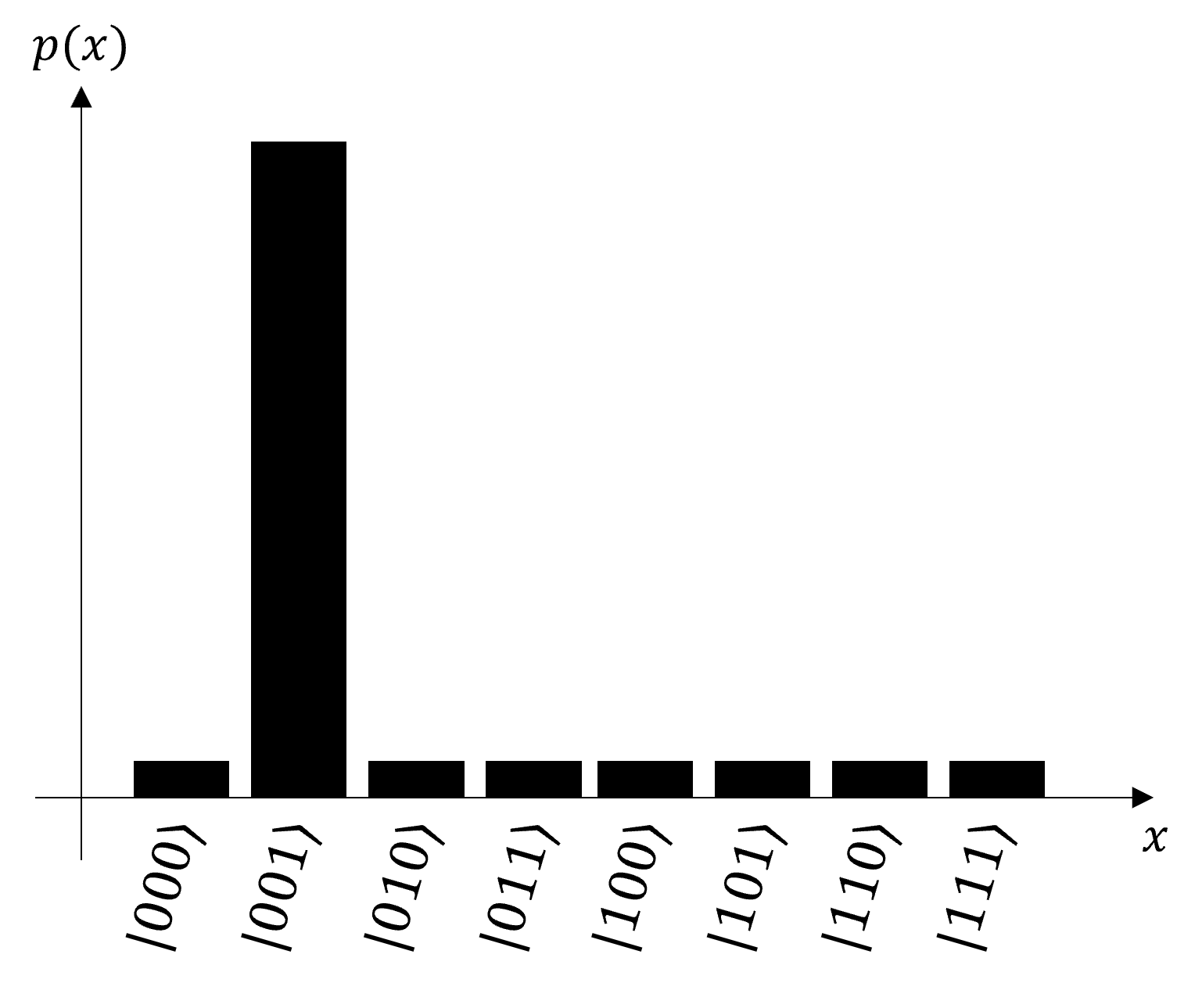}
    \caption{Final distribution for the last iteration. In this case, $f(001) = y_{min} = |0100\rangle$.}
\label{final_dist}
\end{figure}

In this example, the state $|0100\rangle$ will be measured with approximately $100\%$ of probability (See Fig. \ref{final_dist}). The process is iteratively done with the rest of the qubits in order to find the minimum $y_{min} = |0100\rangle$ surely. 

The task of finding a minimum in a vector can be performed using an optimal algorithm, with time complexity $\mathcal{O}(|\Vec{y}|)$. The quantum algorithm proposed in this work solves the same problem on a quantum computer by performing $\mathcal{O}(c\sqrt{\frac{|\Vec{y}|}{t}})$ queries in Grover's oracle, where $c$ is a constant whose value is a number of digits in a binary representation of the initial value and $t$ is the number of marked states. According to complexity theory, a constant doesn't affect time complexity, then it is valid to rewrite the time complexity of this algorithm as $\mathcal{O}(\sqrt{\frac{|\Vec{y}|}{t}})$.

\subsection{Complexity analysis}

\begin{figure} [H]
  \centering  \includegraphics[width=0.4\textwidth]{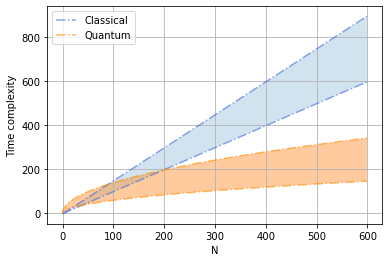}
    \caption{Complexity analysis among algorithms. The shade between lines represents the complexity range among classical and quantum algorithms. The upper and lower bounds of the classical algorithms (blue) have time complexities $\mathcal{O}(\frac{3}{2}N - 2)$ and $\mathcal{O}(N-1)$. For the case of the quantum algorithm (orange), the upper and lower limits were drawn for the cases where $c_q = 14$ and $c_q = 6$, respectively.}
    \label{comp}
\end{figure}

In order to analyze and compare the time complexity between classical and quantum algorithms for different scenarios, we take two classical algorithms with different complexities. For the proposed quantum algorithm, the same was done, but using an increase in complexity by increasing the number of bits in the database values (See Fig. \ref{comp}). We know that classical and quantum algorithms have complexities $\mathcal{O}(c_cN)$ and $\mathcal{O}(c_q\sqrt{N})$, respectively. The constants $c_c$ and $c_q$ indicate, respectively, the constant inherent complexity factor of each classical algorithm and the number of bits of the quantum algorithm's initial guess, as explained in the procedure.

\section{K-means clustering}

\begin{figure} [H]
  \centering  \includegraphics[width=0.4\textwidth]{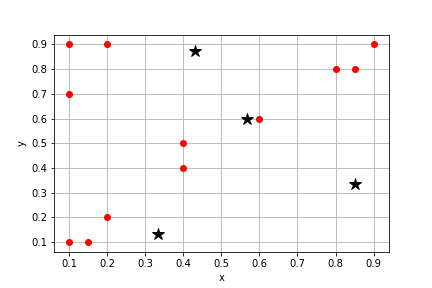}
    \caption{Distribution of $|S|=12$ points (red) in two-dimensional Euclidean space. In this example, four centroids ($K=4$), represented by stars (black), were randomly initialized.}
    \label{random}
\end{figure}

In order to demonstrate the application of minimum search in important computer science tasks, we implemented the clustering algorithm called \textit{K-means}, well known in statistics and unsupervised machine learning. Given a set of points in Euclidean space, the algorithm aims to determine the optimal location of $K$-centroids in order to minimize the sum of all distances between the points and their respective centroids. The objective function can be given by

\begin{equation}
    f(x,y) = \sum_{j=0}^{K}\sum_{i=0}^{|S|}\|p_i^{(j)} - c_i\|^2 
\end{equation}

where $p_i$ is an observed point in Euclidean space, $|S|$ is the total number of observed points, $c_i$ is the position of a centroid and $K$ is the predetermined number of centroids. Fig. \ref{random} shows a distribution of points in the Cartesian plane and the randomly initialized centroids before starting the optimization process. Although a simplified example, this one can be useful to visually demonstrate how the algorithm works.

\begin{figure} [H]
  \centering  \includegraphics[width=0.4\textwidth]{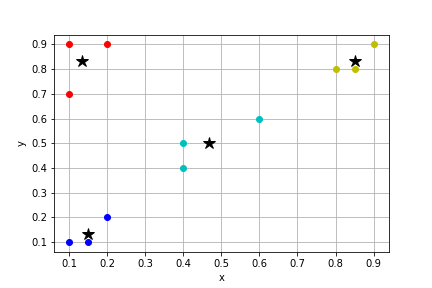}
    \caption{Optimal solution found by the quantum version of \textit{K-means}. Each of the four clusters found by the algorithm is being represented by a color.}
    \label{best}
\end{figure}

We assume that in the quantum version of \textit{K-means}, the distances between each observed point, $p_i$, and all centroids, $S = \{c_0, c_1, c_2, c_3\}$, are stored in QRAM in constant time, that is, $\mathcal{O}(1)$. Clusters are formed at each iteration by the proximity between each point and its closest centroid. The average between the coordinates of each new cluster is calculated and becomes the new centroid. This process is carried out until a certain stopping criterion is satisfied. Fig. \ref{best} shows the best clustering found by the algorithm.

Note that the only difference between this procedure to its classical analog is that the distances between each point and all centroids are stored in a QRAM, and the QMS is used to find the smallest one. Although we are using QMS for a specific example, it can be useful for a huge amount of computational tasks, such as unsupervised machine learning problems.

\section{Conclusions}


Classical computing can be outperformed by quantum computing in a wide range of problems, from those with low to those with high levels of computational complexity. The classical minimum search problem is characterized by linear complexity and is connected to an extensive variety of applications in the domain of computer science. In this scenario, this work proposed a quantum algorithm for finding the minimum value of a database that is quadratically faster than its best classical analogs. The algorithm is based on D\"{u}rr-Hoyer's approach for finding a minimum value in an unstructured list through the use of Grover's algorithm as a subroutine applied to a QRAM that stores values from a defined database. Although it is not considered a complex task, our results show that the suggested QMS algorithm has the potential to significantly reduce the execution time of minimum search algorithms for cases where the database is very large. Moreover, an examination of the complexity of the studied problem was performed in order to highlight the advantages of this quantum algorithm over its classical analogs.
Furthermore, we show how the suggested approach can be used in an unsupervised machine learning task by performing a quantum adaptation of the K-means algorithm. In conclusion, our results demonstrate that it is possible to search for minimums in a classical database by utilizing information stored in a QRAM, which represents a significant contribution to the development of fault-tolerant quantum algorithms.



\begin{acknowledgments}
We thank the Latin American Quantum Computing Center and the High-Performance Computing Center, both from SENAI CIMATEC for supporting this research. We also thank the Quantum Open Source Foundation (QOSF) mentoring program, whose developed project originated this article. 
\end{acknowledgments}

\bibliography{sample}

\end{document}